\documentclass[12pt,a4paper]{article}
\usepackage{amsmath,amssymb}
\usepackage{graphicx}
\usepackage[articletitle=true, maxauthors=40]{achemso}

\usepackage{xcolor}
\usepackage{paralist}

\usepackage{mathtools} 
\usepackage{booktabs}

\usepackage{siunitx}
\usepackage{bm}

\begin{document}

\begin{center}

\vspace*{1cm}

{\LARGE\bf Uncertainty-aware First-principles Exploration of Chemical Reaction Networks\\[2ex]}

\vspace{.5cm}

{\large Moritz Bensberg\footnote{ORCID: 0000-0002-3479-4772}
and Markus Reiher\footnote{email: mreiher@ethz.ch; ORCID: 0000-0002-9508-1565}}\\[2ex]

ETH Z\"urich, Department of Chemistry and Applied Biosciences, Vladimir-Prelog-Weg 2,\\ 8093 Z\"urich, Switzerland\\[4ex]

Date: \hspace*{.31cm} 24.12.2023\\[2ex]

\vspace*{.51cm}

\end{center}

\vspace{.31cm}

\begin{abstract}
Exploring large chemical reaction networks with automated exploration approaches and accurate quantum chemical methods can require prohibitively large computational resources. Here, we present an automated exploration approach that focuses on the kinetically relevant part of the reaction network by interweaving (i) large-scale exploration of chemical reactions, (ii) identification of kinetically relevant parts of the reaction network through microkinetic modeling, (iii) quantification and propagation of uncertainties, and (iv) reaction network refinement. Such an uncertainty-aware exploration of kinetically relevant parts of a reaction network with automated accuracy improvement has not been demonstrated before in a fully quantum mechanical approach.
Uncertainties are identified by local or global sensitivity analysis. The network is refined in a rolling fashion during the exploration. Moreover, the uncertainties are considered during kinetically steering of a rolling reaction network exploration.
We demonstrate our approach for Eschenmoser--Claisen rearrangement reactions. The sensitivity analysis identifies that only a small number of reactions and compounds are essential for describing the kinetics reliably, resulting in efficient explorations without sacrificing accuracy and without requiring prior knowledge about the chemistry unfolding.
\end{abstract}


\newpage

\section{Introduction}\label{sec:intro}
If chemical compounds react in a flask in the laboratory, there will be a large number of reaction paths conceivable, leading to a complex network of elementary reaction steps with potentially many products.
Detailed knowledge of such a reaction network is required for any kind of rational reaction optimization in order to prevent the formation of side products while promoting a desired reaction path.
Constructing such reaction networks is facilitated by automated reaction network exploration protocols based on quantum chemical calculations (see Refs.~\citenum{Dewyer2017, Simm2018, Green2019, Unsleber2020, Maeda2021, Baiardi2021, Steiner2022} for reviews). These protocols construct large reaction networks with automated algorithms, therefore reducing the amount of manual work and the chance of overlooking essential reaction channels compared to manual investigations.
After calculating the free energies for all compounds and rate constants in the network, these networks can be directly subjected to microkinetic modeling to predict products, key intermediates of the reaction, and concentration profiles.

Since the objective of a reaction network exploration is to derive a quantitative high-fidelity model of a chemical reaction in experiment, the emerging chemical reaction network should focus on the chemistry of the reactive system under experimental conditions. This means that the automated exploration must be autonomously steered toward the kinetically relevant part of the network.
To address this challenge, we proposed an automated kinetics-interlaced exploration algorithm\cite{Bensberg2023a} (KIEA) that achieves this through analysis of concentration fluxes obtained from microkinetic modeling during the generation of the network. A related analysis of microkinetic modeling simulations is used in the reaction mechanism generator~\cite{Gao2016, Goldsmith2017, Liu2021, Johnson2022} (RMG), which focuses on combustion chemistry\cite{Matheu2001, Geem2006, Blurock2012}. The algorithm in RMG follows a greedy strategy during the exploration, focusing on an in-depth exploration of single reaction paths~\cite{Johnson2023a} rather than on a broad exploration, as facilitated by KIEA.
Sumiya and Maeda\cite{Sumiya2020} suggested an alternative approach to steer automated explorations by only analyzing the rate constant matrix of the reaction network and avoiding explicit microkinetic modeling. However, their approach is restricted to a single potential energy surface, implying that the atom composition of every compound in the network must be the same.
Apart from these approaches, a shortest-path analysis\cite{Xie2021, Blau2021, Tuertscher2022}, such as provided by Pathfinder\cite{Tuertscher2022}, which takes kinetic information of the reaction network into account, can also quantify how accessible a compound in the reaction network is and, hence, steer the exploration of reaction networks.

All these steering approaches depend crucially on the accuracy of the kinetic and thermodynamic parameters of the underlying reaction network. However, accurate quantum chemical methods require tremendous computational resources, making a large-scale exploration of tens of thousands of reactions challenging, if not impossible. Therefore, a refinement-based strategy is a way out of this problem in which the network is initially explored with a computationally efficient but less reliable method before refining the energies (and sometimes the structures) of the compounds in a second step\cite{Simm2018a, Xu2023a, GarayRuiz2022, Stan2022, Bensberg2023a}.
Because of the high computational cost of accurate quantum chemical calculations, the refinement is generally executed after the exploration and is limited only to a small set of reactions and compounds that dominate the overall kinetics.~\cite{Sutton2015, Bensberg2023a}
These reactions and compounds can be identified by sensitivity analysis of the microkinetic model with respect to its kinetic parameters (\emph{e.g.}, rate constants or free energies) \cite{Meskine2009, Sutton2015, Zhao2017, Proppe2018, Gao2020}.

Since autonomous steering of an automated reaction network exploration should depend on the kinetics of the network, reliable kinetic parameters are crucial during exploration.
Therefore, we propose to explicitly interweave (i) an unfolding exploration of the reaction network with (ii) the identification of kinetically relevant reactions and compounds and (iii) the refinement of the kinetic parameters in one algorithm.

Our algorithm combines KIEA to steer the exploration with
an integrated refinement of structures and energies (IRES) that identifies important reactions and compounds through local one-at-a-time (OAT) or Morris sensitivity analysis\cite{Morris1991} of the microkinetic modeling output. IRES then refines structures, reaction paths, and energies in the network fully automatically.
The Morris sensitivity analysis not only identifies important parameters in the microkinetic model, it also quantifies the uncertainty in the predicted concentrations. We exploit this fact and demonstrate how the uncertainties can be directly included in KIEA.

This work is structured as follows:
First, we develop the IRES algorithm in Section~\ref{sec:concepts}, detailing our microkinetic modeling and sensitivity analysis approaches. In Section~\ref{sec:comp-methods}, we provide technical details and introduce the Eschenmoser--Claisen reaction, which serves as an example for developing our exploration approach. We then demonstrate the IRES-KIEA in Section~\ref{sec:results} and conclude in Section~\ref{sec:conclusion}.

\section{Conceptual Considerations\label{sec:concepts}}

\subsection*{Microkinetic Modeling\label{sec:micro-kinetic_modeling}}
For microkinetic modeling, the ordinary differential equations describing the mass-action kinetics of a chemical reaction network are integrated to obtain the concentration trajectories $c_n(t)$ for each species $n$.
The forward $(+)$ and backward $(-)$ reaction rates $f^{+/-}_I$ of the reaction $I$ are given as
\begin{align}
    f^{+/-}_I = k_I^{+/-} \prod_{n} [c_n (t)]^{S^{+/-}_{nI}}~,
\end{align}
with the forward and backward reaction rate constants $k^+_I$ and $k^-_I$, respectively, and the stoichiometric coefficients $S^{+/-}_{nI}$ of the species in reaction $I$
Accordingly, the differential equation describing the change of concentration of species $n$ is given by
\begin{align}
    \frac{\mathrm{d} c_n(t)}{\mathrm{d} t} =  \sum_I \left\{ \left[S^{-}_{nI} - S^{+}_{nI} \right] \left[ f^{+}_I - f^{-}_I \right]  \right\}~,
\end{align}
and the total concentration flux passing through reaction $I$ by
\begin{align}
    F_I = \int_{t_0}^{t_\mathrm{max}} |f^{+}_I - f^{-}_I|~,
\end{align}
where $t_0$ and $t_\mathrm{max}$ denote the start and end times of the microkinetic modeling simulation, respectively.
The concentration flux passing through species $n$ thus reads
\begin{align}
    c^\mathrm{flux}_n = \sum_I \left(S^{-}_{nI} + S^{+}_{nI}\right) F_I~.
\end{align}

We approximate the reaction rate constants $k_I^+$ by Eyring's absolute rate theory~\cite{Eyring1935, Truhlar1996}
\begin{align}
    k_I^+ = \Gamma \frac{k_B T}{h} \exp\left[-\frac{\Delta G^\ddagger_I}{k_B T}\right]~,
    \label{eq:forward_rate_constant}
\end{align}
where $G^\ddagger_I$ is the free energy of activation of reaction $I$, $h$ is Planck's constant, $T$ the temperature, $k_B$ Boltzmann's constant, and $\Gamma$ the transmission coefficient (assumed to be $\Gamma = 1$ in the following).
To ensure that the reaction is thermodynamically balanced, the reverse rate constant $k_I^-$ is then expressed with the equilibrium constant $K_I$ as
\begin{align}
    k_I^- = k_I^+ / K_I~.
    \label{eq:backward_rate_constant}
\end{align}
The equilibrium constant $K_I$ is defined as usual
\begin{align}
    K_I = \exp\left[-\frac{\Delta G_I}{k_B T} \right]
\end{align}
with the free energies $G_n$ of the species on the reaction's right-hand side (RHS) and left-hand side (LHS):
\begin{align}
    \Delta G_I = \sum_{n\in \mathrm{RHS}(I)} G_n - \sum_{m\in \mathrm{LHS}(I)} G_m
    \label{eq:free_energy_difference}
\end{align}

\subsection*{Sensitivity Analysis\label{sec:sensitivity-analysis}}

The calculation of the parameters $G_n$ and $\Delta G^\ddagger_I$ required for the microkinetic modeling [see Eqs.~(\ref{eq:forward_rate_constant}) and (\ref{eq:free_energy_difference})] will always be subject to various approximations, leading to an uncertain microkinetic modeling output. To reduce the uncertainty in the microkinetic modeling output, our IRES approach identifies the most influential parameters ($G_n$ and $\Delta G^\ddagger_I$) through sensitivity analysis and refines them by carrying out more accurate calculations in a fully autonomous fashion. The objective of IRES is to increase the accuracy of the continued reaction network exploration driven by KIEA, which relies on the concentration fluxes $c^\mathrm{flux}_n$ and maximum concentrations $c^\mathrm{max}_n$ encountered during the microkinetic modeling.
Because $c^\mathrm{max}_n$ is a lower bound for $c^\mathrm{flux}_n$ for compounds with zero starting concentration, it is the key output of the microkinetic modeling simulation and, therefore, analyzed by sensitivity analysis.

In local OAT sensitivity analysis, the relevance of an input parameter on the model output is calculated by changing one input parameter $x_i$ at a time from the baseline parameters $\pmb{X}_\mathrm{base}$ (such as the most accurate free energies available) and evaluating the model output. Therefore, only one parameter differs from the baseline parameters during model evaluation. To provide an upper limit for the error of $c^\mathrm{max}_n$, the maximum effect of the parameter uncertainty on $c^\mathrm{max}_n$ is crucial. For realistic variations of the parameters, we vary the free energies in the microkinetic modeling within their uncertainty bounds. We can expect the effect of this variation to be the largest if we change the parameter by its uncertainty, \emph{i.e.}, to the edge of the range of likely values. Therefore, we define the modification of the input parameters as
\begin{align}
\begin{split}
    x_i &\rightarrow x^\mathrm{u}_i = x_i + u(x_i) \\
    x_i &\rightarrow x^\mathrm{l}_i = \begin{cases}
                        x_i - u(x_i) &\text{if } x_i \text{ is a free energy } G_n\\
                        \max\left(0.0, x_i - u(x_i)\right) &\text{if } x_i \text{ is a free energy of activation } \Delta G^\ddagger_I,
                     \end{cases}
\end{split}
\label{eq:parameter_values}
\end{align}
where $u(x_i)$ is the uncertainty we expect for parameter $x_i$, and $x^\mathrm{u}_i$ and $x^\mathrm{l}_i$ denote the most extreme upper and lower parameter values of $i$, respectively. Care must be taken when modifying the free energies to avoid negative backward barriers. In such cases, the forward reaction barrier is increased to give a zero backward barrier.

To derive a sensitivity measure $\delta c^\mathrm{max}_i$, we collect the maximum concentrations $c^\mathrm{max}_n(\pmb{X}_i^\mathrm{l/u})$ from the OAT model evaluations and calculate their absolute maximum change
\begin{align}
    \delta c^\mathrm{max}_i = \max_{\mathrm{l,u}}\max_{n}\left| c^\mathrm{max}_n(\pmb{X}_\mathrm{base}) - c^\mathrm{max}_n(\pmb{X}_i^\mathrm{l/u}) \right|
\end{align}
compared to the baseline model's maximum concentrations $c^\mathrm{max}_n(\pmb{X}_\mathrm{base})$, where $\pmb{X}_i^\mathrm{l/u} = (x_1, \ldots, x_{i-1}, x_i^\mathrm{l/u}, x_{i+1}, \ldots x_k)$ are the modified parameters from the OAT procedure and $k$ is the total number of parameters.

Because KIEA disregards any compound with negligible concentration flux in following microkinetic modeling steps\cite{Bensberg2023a}, refinement of these compounds cannot affect the exploration. Therefore, the sensitivity analysis can be accelerated by (i) varying free energies only if the associated species shows a concentration flux $c^\mathrm{flux}_n > \tau_\mathrm{flux}^\mathrm{kin}$, and (ii) by varying free energies of activation only if the reaction exhibits a flux $F_I >  \tau_\mathrm{flux}^\mathrm{kin}$.

The baseline parameters $\pmb{X}_\mathrm{base}$ can be understood as one point in the possible input space given by all possible values within the input's uncertainty. Because local OAT sensitivity analysis samples only a tiny part of this input space close to the baseline point, it is often criticized for being unreliable in identifying essential model parameters and may fail to provide the correct picture of the sensitivities and model output uncertainties\cite{Saltelli2005, Saltelli2010}.

A computationally affordable alternative to local sensitivity analysis is Morris sensitivity analysis\cite{Morris1991}, where a grid of equally spaced input values is formed for each parameter from the range of possible values. This range is given as the interval between the values of $x_i$ in Eq.~(\ref{eq:parameter_values}). Afterward, the model is evaluated for a set of $N$ samples $\pmb{X}_r = (x_{r,0}, ... x_{r,k})$, drawn at random from initially selected parameter values, where $x_{r,0},...x_{r,k}$ are the $k$ model parameters for sample $r$.
Then, each parameter value of $\pmb{X}_r$ is changed one-at-a-time in random order to a neighboring value $x^\prime_{r, i}$ on the parameter grid. The parameters $x^\prime_{r, i}$ are not returned to their initial values $x_{r, i}$. Therefore, this algorithm creates a trajectory $\pmb{\Tilde{X}}_r$ through the input space starting at $\pmb{X}_r$. By this procedure, Morris sensitivity analysis covers a significantly larger part of the input space than local OAT analysis. It is able to identify crucial parameters in the model with a relatively small number of samples $N$, typically in the range between 10 and 20\cite{Saltelli2005}.

To quantify the maximum effect of an input parameter on the maximum concentrations, we define a sensitivity measure as 
\begin{align}
    \mu^{*\mathrm{max}}_{i} = \max_n \mu^*_{ni}~,
\end{align}
where $\mu^*_{ni}$ is the expectation value of the absolute elementary effect\cite{Campolongo2007} for parameter $i$, and maximum concentration $c^\mathrm{max}_n(\pmb{X}_r)$
\begin{align}
    \mu^*_{ni} = \left\langle \frac{\left|c_n^\mathrm{max}(\Tilde{x}_{r, 1}, \ldots \Tilde{x}_{r, i-1}, x_{r, i} + \Delta, \Tilde{x}_{r, i+1}, \ldots \Tilde{x}_{r, k}) - c_n^\mathrm{max}(\pmb{X}_r)\right|}{\Delta} \right\rangle_r~.
\end{align}
Here, $\Delta$ is the difference between the values for parameter $i$ on its parameter grid, and the tilde (\emph{i.e.}, $\Tilde{x}_{r, j}$ instead of $x_{r, j}$) highlights that these parameters may have been changed before because of the random order in the parameter modification during the sensitivity analysis.

Since Morris sensitivity analysis provides an adequate sampling of the input parameter space, the spread in the microkinetic modeling output provides an uncertainty measure for the concentrations.
This allows us to define an uncertainty-aware version of KIEA.
Instead of exploring unimolecular and bimolecular reactions based on the criteria $c_n^\mathrm{flux} > \tau_\mathrm{flux}$ and $c_n^\mathrm{max} c_m^\mathrm{max} > \tau_\mathrm{max}$, respectively\cite{Bensberg2023a}, we include the concentration spread by reformulating these criteria as
\begin{align}
    \bar{c}^\mathrm{flux}_n + \sigma(c^\mathrm{flux}_n) > \tau_\mathrm{flux}
    \label{eq:flux-cond-sigma}
\end{align}
and
\begin{align}
    [\bar{c}^\mathrm{max}_n + \sigma(c^\mathrm{max}_n)] [\bar{c}^\mathrm{max}_n + \sigma(c^\mathrm{max}_n)] > \tau_\mathrm{max}~.
    \label{eq:max-cond-sigma}
\end{align}
Here, $\bar{c}^\mathrm{flux}_n$ and $\sigma(c^\mathrm{flux}_n)$ are arithmetic mean and standard deviation of the concentration flux of compound $n$, and $\bar{c}^\mathrm{max}_n$ and $\sigma(c^\mathrm{max}_n)$ are arithmetic mean and standard deviation of the compound's maximum concentration, respectively. Mean and standard deviation are calculated over the ensemble of microkinetic modeling simulations in the Morris sensitivity analysis.

\section{Computational Methodology\label{sec:comp-methods}}

\subsection*{The Eschenmoser--Claisen Rearrangement\label{sec:example}}
\begin{figure}
    \centering
    \includegraphics[width=\textwidth]{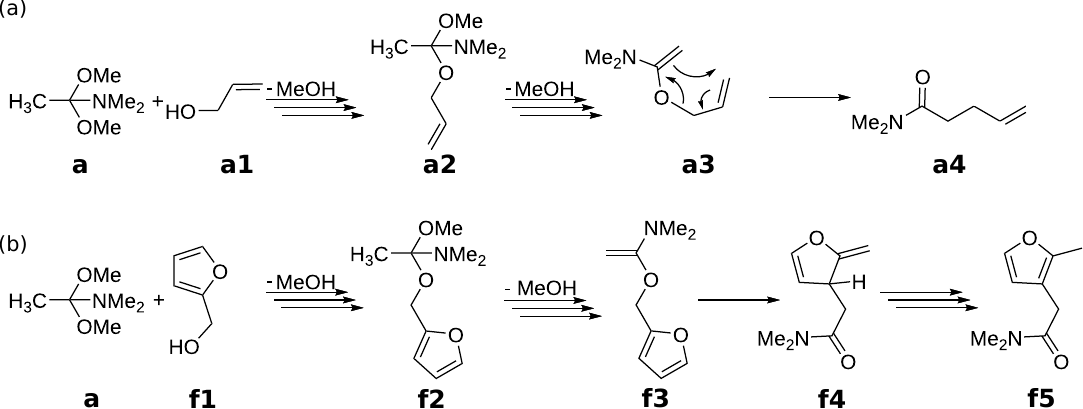}
    \caption{Sketch of the reaction mechanisms of the Eschenmoser--Claisen rearrangement reactions of allyl alcohol (a) and furfuryl alcohol (b). The notation with multiple arrows indicates that the reaction may not be a single elementary reaction step.}
    \label{fig:mechanism_sketch}
\end{figure}

To demonstrate our IRES-KIEA approach, we chose the Eschenmoser--Claisen rearrangement\cite{Wick1964} of allyl alcohol \textbf{a1} and of furfuryl alcohol \textbf{f1}.
The rearrangement of furfuryl alcohol was first reported in 1969\cite{Felix1969} in dimethylformamide at $160~\si{\celsius}$ after $24~\si{h}$. However, there is no experimental report on the Eschenmoser--Claisen rearrangement of allyl alcohol.
Still, allyl alcohol represents the main reactive moiety in the reaction, making it an ideal model reactant for a general Eschenmoser--Claisen rearrangement. A sketch of the reaction mechanisms is shown in Fig.~\ref{fig:mechanism_sketch}.
The elevated reaction temperature is required for the rate-limiting initial alcohol exchange and methanol elimination to form the intermediates \textbf{a3} and \textbf{f3}, respectively,  before the Claisen rearrangement step occurs\cite{Castro2004}.
In the case of the furfuryl-based rearrangement [Fig.~\ref{fig:mechanism_sketch}(b)], the product of the Claisen-rearrangement \textbf{(f4)} step undergoes an H-shift to
re-establish aromaticity in the furan moiety and form the final product \textbf{f5}.
The Eschenmoser--Claisen rearrangement reaction is an $E$ stereo-selective, [3, 3] sigmatropic rearrangement of allyl alcohols and N, N-dimethylacetamide-dimethyl acetal \textbf{a} at reduced temperatures of around $150~\si{\celsius}$
compared to other Claisen-type rearrangements\cite{Kurti2005}.
The reaction is employed in natural product synthesis because of the mild reaction conditions and its stereoselectivity\cite{Daniewski1992, Chen1993, Williams2000, Loh2000}.

\subsection*{Reaction Network Exploration}
In our \textsc{Scine} software framework\cite{Scine}, reaction networks are encoded in terms of \emph{structures}, which are local minima on Born-Oppenheimer potential energy surfaces, and \emph{elementary steps}, which represent transitions between
local minima on a potential energy surface\cite{Unsleber2020}. These transitions proceed either through a transition state or are barrier-less processes (\emph{e.g.}, in the case of the association of two molecules to form a weakly
interacting complex). Several structures (typically conformers) are grouped into \emph{compounds} according to their charge, spin multiplicity, and the abstract molecular graph and structure representation determined by our software module \textsc{Molassembler}\cite{Sobez2020, Bensberg2023g}. A structure containing multiple molecules is grouped into so-called \emph{flasks}, in which reactive complexes are formed. Elementary steps are grouped into \emph{reactions} so that compounds or flasks associated with the structures that are connected by the elementary steps can be related.

\subsection*{Microkinetic Modeling}
The mass-action kinetics were integrated at the level of compounds and flasks as kinetic species and reactions describing the transition between these species.
Because we did not perform exhaustive conformer searches for every compound, flask, and transition state, we approximated $G_n$ by the minimum of the harmonic-oscillator/particle-in-a-box/static-rotor free energy approximation $G^\mathrm{HPS}_{i}$ calculated for any structure $i$ of the compound or flask $n$
\begin{align}
    G_n &= \min_{i \in n} G^\mathrm{HPS}_{i}~,
    \label{eq:free_energy}
\end{align}
where $G^\mathrm{HPS}_{ni}$ is given by
\begin{align}
    G^\mathrm{HPS}_{i} = E^\mathrm{elec}_i + \delta G^\mathrm{vib}_i + \delta G^\mathrm{rot}_i + \delta G^\mathrm{trans}_i + \delta G^\mathrm{solv}~.
\end{align}
Here, $E^\mathrm{elec}_i$, $\delta G^\mathrm{vib}_i$, $\delta G^\mathrm{rot}_i$, $\delta G^\mathrm{trans}_i$, $\delta G^\mathrm{solv}$ are the electronic energy, the harmonic vibrational free energy correction, the free energy correction from the static rotor model, the translational free energy correction from the particle-in-a-box model, and the solvation free energy correction, respectively.
We calculated the translational free energy contribution for a concentration of $1.0~\si{mol.L^{-1}}$ to account for the typical standard state free energy correction in solution\cite{BenNaim1987}.

Similar to Eq.~(\ref{eq:free_energy}), we calculated the free energies of activation $\Delta G_I^\ddagger$ as
\begin{align}
    \Delta G_I^\ddagger&= \min_{i \in I}( G^\mathrm{HPS}_{i}) - \sum_{n\in \mathrm{LHS}(I)} G_n\\
                       &= G^\ddagger_I - \sum_{n\in \mathrm{LHS}(I)} G_n~,
\end{align}
\emph{i.e.}, as the difference between the minimal $G^\mathrm{HPS}_{i}$ approximation for a transition state of the reaction and LHS's free energy.
In the case of barrier-less reactions, where transition states are not available, the free energy approximation for the transition state $G^\ddagger_I = \min_{i \in I}( G^\mathrm{HPS}_{Ii})$ was replaced by the maximum of the free energies of RHS and LHS
\begin{align}
 G^\ddagger_I = \max\left(\sum_{n\in \mathrm{LHS}(I)} G_n, \sum_{n\in \mathrm{RHS}(I)} G_n\right)~.   
\end{align}

\subsection*{Electronic Structure Models\label{sec:electronic_structure}}

To reduce the number of single-point calculations, we refined electronic energies by coupled cluster calculations\cite{Bartlett2007, Shavitt2009} only for a subset of structures.
Either the structures were discovered as part of an elementary step with a barrier lower than $250.0~\si{kJ.mol^{-1}}$ or during the sensitivity-based refinement.

To maximize the efficiency of the exploration and achieve sufficient accuracy for the microkinetic modeling, we calculated the electronic energy contribution to the free energy with a different electronic structure method than employed for the reaction exploration, structure optimization, and harmonic frequency calculations. These model combinations will be denoted as \texttt{electronic energy model//structure optimization and frequency model}.
We applied the following three ranks for our refinement-based exploration strategy:
\begin{itemize}
    \item[(I)] PBE0-D3//GFN2-xTB
    \item[(II)] PBE0-D3//PBE-D3
    \item[(III)] DLPNO-CCSD(T)//PBE-D3 
\end{itemize}
Here, we denote the exchange--correlation hybrid functional by Adamo and Barone\cite{Adamo1999} as PBE0 and the pure functional by Perdew, Burke, and Ernzerhof's as PBE\cite{Perdew96}. Both functionals were corrected for long-range dispersion by Grimme's D3 correction\cite{Grimme2010}, including Becke--Johnson damping\cite{Grimme2011}.
GFN2-xTB denotes the semi-empirical tight binding method developed by Bannwarth \emph{et al.}\cite{Bannwarth2019}, and DLPNO-CCSD(T) refers to domain-based local pair natural orbital coupled cluster with singles, doubles, and perturbative triples excitations\cite{Riplinger2013a, Riplinger2016} with tight pair natural orbital (PNO) thresholds.
PBE0-D3 and DLPNO-CCSD(T) calculations were carried out with the def2-TZVP basis set\cite{Ahlrich2005} and PBE-D3 calculations with the def2-SV(P) basis set\cite{Ahlrich2005}.
Furthermore, the conductor-like screening model\cite{Klamt1993} represented the solvent in the DFT calculations (dielectric constants ($\epsilon$), and solvent radii ($r_\mathrm{solv}$): toluene: $\epsilon=2.38$, $r_\mathrm{solv}=3.48~\si{au}.$, acetonitrile: $\epsilon=37.5$, $r_\mathrm{solv}=2.76~\si{au}.$), whereas the generalized Born and surface area\cite{Onufriev2004, Sigalov2006} model described the solvent in the GFN2-xTB calculations, and the conductor-like polarizable continuum model\cite{Barone1998} (toluene: $\epsilon=2.4$, $r_\mathrm{solv}=1.3~\si{au}.$, acetonitrile: $\epsilon=36.6$, $r_\mathrm{solv}=1.3~\si{au}.$) represented the solvent in the DLPNO-CCSD(T) calculations.

Free energies $G_n$ for the microkinetic modeling were calculated according to the first two ranks of the electronic structure model hierarchy, \emph{i.e.}, the electronic energies were always calculated with PBE0-D3 to ensure comparable energies. The hierarchy was implemented as follows: If the free energy calculated with PBE0-D3//PBE-D3 was available in the database, it was preferred over a PBE0-D3//GFN2-xTB free energy approximation.
The free energies of activation $\Delta G^\ddagger_I$ were calculated similarly, including all three hierarchy ranks. 

All IRES-based explorations were performed with PBE0-D3//GFN2-xTB as the initial electronic structure method. During the exploration, free energies found to be important by the sensitivity analysis were refined with PBE0-D3//PBE-D3 and free energies of activation with DLPNO-CCSD(T)//PBE-D3.

The initial transition state  GFN2-xTB structures were refined by double-ended reaction path optimizations with PBE-D3 (basis set and solvent models as detailed above) for the ten energetically most favorable elementary steps within $20.0~\si{kJ.mol^{-1}}$ of the lowest PBE0-D3//GFN2-xTB free energy transition state, as described in Ref.~\citenum{Bensberg2023a}. In this double-ended reaction path optimization, the minimum energy path is obtained by curve optimization\cite{Vaucher2018}, the transition state is optimized, and the reactants and reaction products are obtained from an intrinsic reaction coordinate scan.
Then, we calculated the electronic energy for each newly optimized stationary point with DLPNO-CCSD(T) and the vibrational harmonic frequencies with PBE-D3.
To increase the number of successfully refined reactions, we restarted any unsuccessful transition state optimization with a lowered trust radius (to $0.05~\si{Bohr}$ instead of the original $0.1~\si{Bohr}$) and increased the maximum number of iterations (to 250 instead of the original 100).
The accuracy of the free energies of the compounds and flasks was increased by optimizing the ten structures with the lowest value of $G_i^\mathrm{HPS}$ with PBE-D3. These structures were chosen to be at most $20.0~\si{kJ.mol^{-1}}$ higher in energy (PBE0-D3//GFN2-xTB) than the most stable structure. Then, PBE0-D3 electronic energies were calculated for the re-optimized structures, and the vibrational harmonic frequencies were calculated with PBE-D3.

\subsection*{Exploration Protocols}\label{sec:ansatz}
The reaction network was explored with the programs of the \textsc{SCINE} software suite. \textsc{Chemoton}\cite{Unsleber2022, Bensberg2022i} was employed to sort structures and elementary steps and create the input for the individual electronic structure calculations. The exploration calculations were then performed by \textsc{Puffin}\cite{Bensberg2023f} and \textsc{ReaDuct}\cite{Brunken2022, Vaucher2018, Bensberg2023e}.
The electronic structure calculations were performed by external programs: Electronic energies and nuclear gradients were provided by \textsc{Turbomole}\cite{Ahlrichs1989, turbomole741} (version 7.4.1) and \textsc{xTB}\cite{Bannwarth2020} (version 6.5.1) for all DFT models and for GFN2-xTB, respectively. The DLPNO-CCSD(T) electronic energies were calculated with \textsc{Orca}\cite{Neese2022} (version 5.0.2).

Specific reaction conditions for the Eschenmoser--Claisen rearrangement of allyl alcohol were not reported in the literature. We assumed a temperature of $150~\si{\celsius}$ and toluene as a solvent for our exploration because these conditions are close to the conditions reported in the original publication of the Eschenmoser--Claisen rearrangement\cite{Wick1964} and for Eschenmoser--Caisen rearrangements in general\cite{Kurti2005}.
Furthermore, the reaction network of the rearrangement of furfuryl alcohol was explored at $160~\si{\celsius}$ and acetonitrile as a solvent instead of dimethylformamide as reported in Ref.~\citenum{Felix1969}. Acetonitrile was assigned a dielectric constant of $\epsilon = 37.5$, which is similar to that of dimethylformamide ($\epsilon = 37$), but, in contrast to dimethylformamide, solvent parameters were available for all electronic structure methods employed.

We explored the reaction networks of both reactions combined with the local OAT sensitivities for IRES-KIEA with the thresholds $\tau_\mathrm{max} = 1\cdot 10^{-3}~\si{mol^{2}.L^{-2}}$ and $\tau_\mathrm{flux} = 1\cdot 10^{-2}~\si{mol.L^{-1}}$ to select compounds for the exploration of bimolecular and unimolecular reactions, respectively. The maximum time for the microkinetic modeling simulations was set to $t_\mathrm{max} = 24~\si{h}$ to match the experimental reaction conditions. We set the starting concentrations for both reactants to $1~\si{mol.L^{-1}}$ to avoid biasing the exploration to unimolecular kinetics of \textbf{a} (note that \textbf{a} is commonly used in excess of $1.3$ (Ref.~\citenum{Wick1964}) to $2$ (Ref.~\citenum{Felix1969}) equivalents in the experiment).

For comparison, we explored the reaction network of the Eschenmoser--Claisen rearrangement of allyl alcohol with PBE0-D3//GFN2-xTB and DLPNO-CCSD(T)//PBE-D3 with the same KIEA settings as in the local OAT-based explorations. Note that we calculated the free energies for the microkinetic modeling in the DLPNO-CCSD(T)//PBE-D3 exploration with PBE0-D3//PBE-D3 and only the free energies of activation with DLPNO-CCSD(T)//PBE-D3.

The sensitivity measures $\delta c^\mathrm{max}_i$ were calculated after each microkinetic modeling simulation in KIEA with a truncation threshold of $\tau_\mathrm{flux}^\mathrm{kin} = 1\cdot 10^{-5}~\si{mol.L^{-1}}$. Refinement calculations were started for reactions, compounds, and flasks if $\delta c^\mathrm{max}_i > 1\cdot 10^{-2}~\si{mol.L^{-1}}$ for their associated free energy of activation or free energy parameter $i$. We chose a threshold of $1\cdot 10^{-2}~\si{mol.L^{-1}}$ for the maximum concentration change to match the threshold $\tau_\mathrm{flux}$, as this choice reduced the uncertainty in $c_n^\mathrm{flux}$ and $c_n^\mathrm{max}$ for compounds that are either significantly populated during the exploration or at the edge of being explored further by KIEA.

In addition to the local OAT-based IRES strategy, we explored both Eschenmoser--Claisen reactions with the uncertainty-aware algorithm based on Morris sensitivity analysis and the KIEA exploration conditions given in Eqs.~(\ref{eq:flux-cond-sigma}) and (\ref{eq:max-cond-sigma}). The Morris sensitivity indices were calculated with four levels in the parameter grid and $N=20$ samples.
This definition of the exploration criteria in Eqs.~(\ref{eq:flux-cond-sigma}) and (\ref{eq:max-cond-sigma}) explicitly includes a measure of the uncertainty of the maximum concentrations and concentration fluxes through their standard deviation. Therefore, we chose the thresholds $\tau_\mathrm{max}= 1\cdot 10^{-2}~\si{mol^{2}.L^{-2}}$ and $\tau_\mathrm{flux}= 1\cdot 10^{-1}~\si{mol.L^{-1}}$ significantly higher than in the local OAT-based explorations.
We refined parameters $i$ if $\mu^{*\mathrm{max}}_{i}$ exceeded a refinement threshold $\tau_\mathrm{ref} = 5\cdot 10^{-2}~\si{mol.L^{-1}}$, which means that a small modification of the parameter is expected to change at least one maximum concentration by $5\cdot 10^{-2}~\si{mol.L^{-1}}$. Similar to the threshold choice for the local OAT sensitivities (\emph{vide supra}), we chose the value of $\tau_\mathrm{ref}$ such that it was close to $\tau_\mathrm{flux}$, therefore reducing the uncertainty in concentration fluxes and maximum concentrations for compounds which were close to being considered for further exploration.

For the Morris sensitivity analysis, we select the sampling trajectories $\pmb{\Tilde{X}}_r$ of the microkinetic models with up to 1000 parameters (elements of $\pmb{X}_r$, that is free energies $G_n$ or activation energies $\Delta G_I^\ddagger$) through a variant of Morris sensitivity analysis proposed by Saltelli and coworkers\cite{Campolongo2007} that maximizes the input space covered by the sensitivity analysis, instead of relying on an initially small number of random points, as discussed in Section~\ref{sec:sensitivity-analysis}. This modified Morris approach became prohibitively slow for large microkinetic models with more than 1000 parameers, for which we relied on random trajectories, as proposed originally by Morris\cite{Morris1991}. Furthermore, we applied a variant of the flux-based screening procedure from the local OAT sensitivities in the case of microkinetic models with more than 1000 parameters. In such cases, we restricted the Morris sensitivity analysis to parameters associated with compounds and flasks with $c^\mathrm{flux}_n > 1\cdot 10^{-9}~\si{mol.L^{-1}}$, and reactions with $F_I > 1\cdot 10^{-9}~\si{mol.L^{-1}}$ in the baseline microkinetic modeling simulation. We chose this screening procedure as a compromise to prevent the tens of thousands of microkinetic model evaluations from becoming the bottleneck of the exploration. We chose the screening threshold as $1\cdot 10^{-9}~\si{mol.L^{-1}}$, and hence, significantly lower than for the local OAT sensitivities. Note that our uncertainty-aware exploration protocol also considers the variance in the concentration flux, which is only available after the sensitivity analysis.
The Morris sensitivity analysis and sampling were performed through an interface to the \textsc{Sensitivity Analysis Library}\cite{Herman2017, Iwanaga2022}.
All microkinetic modeling simulations in this work were executed by an interface to the program Reaction Mechanism Simulator~\cite{Liu2021, Johnson2023}.

\subsection*{Elementary Step Searches}
The reaction network exploration was based on single-ended reaction trial calculations run with the second-generation Newton-trajectory-type algorithm detailed in Ref.~\citenum{Unsleber2022}. For these calculations, the number of bond modifications was limited to two, with at least one intermolecular bond formation for bimolecular reactions.
Furthermore, the reaction trials were restricted by a set of element-specific rules that were chosen to reflect the general textbook-known reactivity of functional groups involved in the mechanism:
\begin{itemize}
    \item Oxygen and nitrogen atoms were always considered reactive.
    \item Hydrogen atoms were considered reactive if part of an ammonium group or at a distance of two bonds to an sp$^2$-hybridized carbon atom or acetal group.
    \item Carbon atoms were considered reactive if sp$^2$-hybridized or neighbors of an sp$^2$-hybridized carbon atom.
\end{itemize}
Furthermore, reaction coordinates were restricted in such a way that they always involved different polarized atoms in bond formation and breaking processes. Atoms were assigned positive and negative polarization identifiers according to their Pauling electronegativities, as described in Ref.~\citenum{Bensberg2023a}.
Moreover, we always assigned positive identifiers to hydrogen atoms and both positive and negative identifiers to sp$^2$-hybridized carbon atoms.

\subsection*{Uncertainty Estimates}

For both sensitivity analysis approaches considered in this work, we required estimates for the uncertainties of $G_n$ and $\Delta G^\ddagger_I$ for PBE0-D3//GFN2-xTB, PBE0-D3//PBE-D3, and DLPNO-CCSD(T)//PBE-D3.
For this, we compared the reaction networks for the Eschenmoser--Claisen rearrangement explored with PBE0-D3//GFN2-xTB and DLPNO-CCSD(T)//PBE-D3 by matching flasks, compounds, and reactions that are accessible from the starting compounds by crossing reaction barriers of less than $400.0~\si{kJ.mol^{-1}}$. We then calculated the differences $\Delta G_n$ of the free energies
\begin{align}
    \Delta G_n &= G_n(\text{PBE0-D3//PBE-D3}) - G_n(\text{PBE0-D3//GFN2-xTB})\label{eq:free-energy-error}
\end{align}
and the differences $\Delta\Delta G_I^\ddagger$ of the activation free energies
\begin{align}
    \Delta\Delta G_I^\ddagger &= \Delta G_I^\ddagger(\text{DLPNO-CCSD(T)//PBE-D3}) - \Delta G_I^\ddagger(\text{PBE0-D3//GFN2-xTB})~.
    \label{eq:barrier-error}
\end{align}
Note that we calculated $\Delta\Delta G_I^\ddagger$ for forward and backward reactions whereas the $\Delta G^\ddagger_I$ parameters in the microkinetic modeling are defined with respect to the LHS of the reaction.

\begin{figure}
    \centering
    \includegraphics[width=\textwidth]{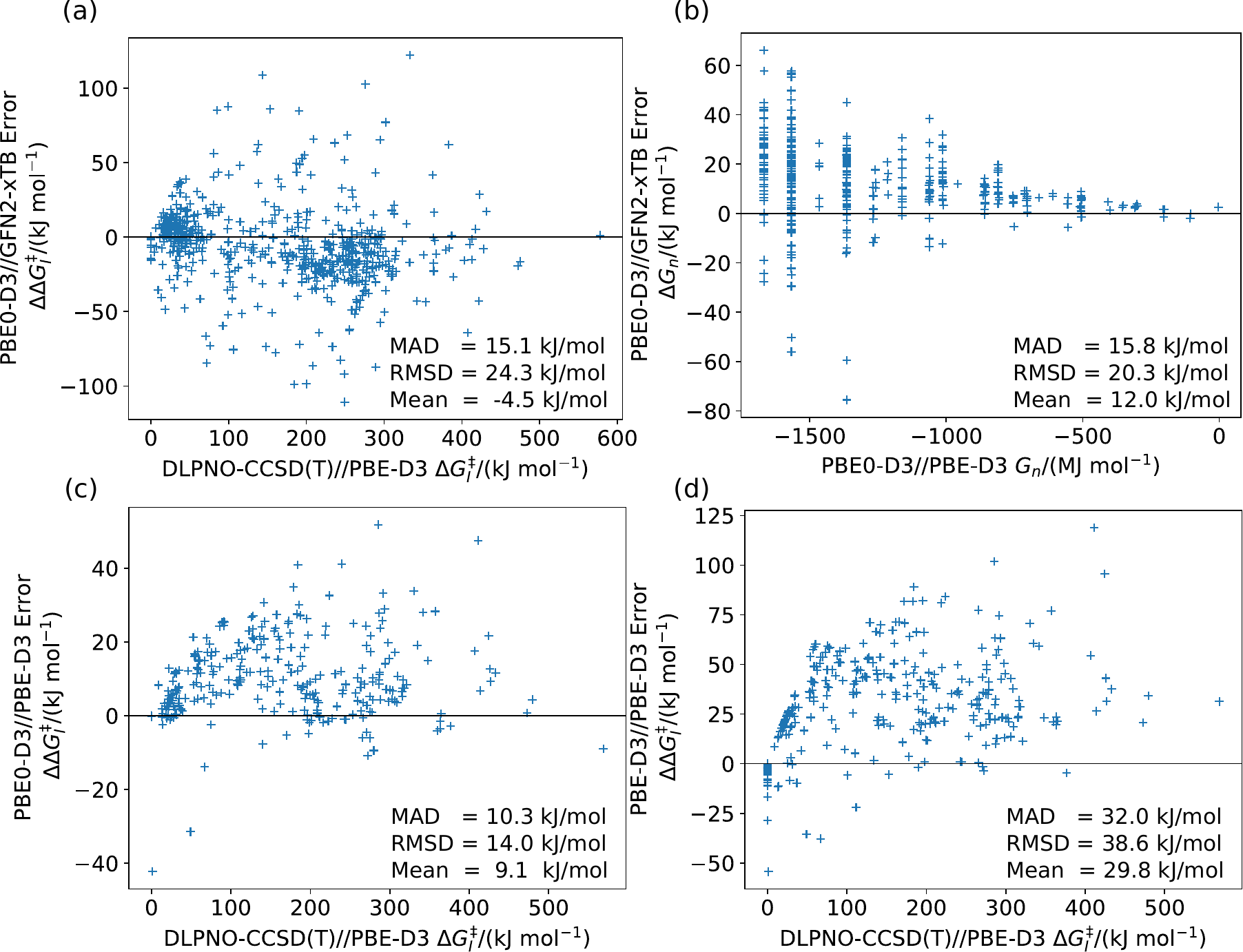}
    \caption{(a, c, d) Errors of the activation energies $\Delta G^\ddagger_I$ calculated with PBE0-D3//GFN2-xTB, PBE0-D3//PBE-D3, and PBE-D3//PBE-D3 with respect to the DLPNO-CCSD(T)//PBE-D3 activation energy. (b) Errors of the free energies $G_n$ calculated with  PBE0-D3//GFN2-xTB with respect to the PBE0-D3//PBE-D3 free energies.}
    \label{fig:gfn2-error-plots}
\end{figure}

The differences $\Delta\Delta G_I^\ddagger$ and $\Delta G_n$ [see Eqs.~(\ref{eq:barrier-error}) and (\ref{eq:free-energy-error})] are shown in Fig.~\ref{fig:gfn2-error-plots} as a function of their reference values $\Delta G_I^\ddagger(\text{DLPNO-CCSD(T)//PBE-D3})$ and $G_n(\text{PBE0-D3//PBE-D3})$, respectively.
The $\Delta\Delta G_I^\ddagger$ values [Fig.~\ref{fig:gfn2-error-plots}(a)] are scattered and can reach the order of magnitude of their reference values in some cases. Nevertheless, the mean absolute difference (MAD) is only $15.1~\si{kJ.mol^{-1}}$, largely because of the high number of relatively low-barrier ($\Delta G_I^\ddagger < 100.0~\si{kJ.mol^{-1}}$) and barrier-less reactions. Furthermore, the mean of $\Delta\Delta G_I^\ddagger$ is $-4.5~\si{kJ.mol^{-1}}$, which suggests that PBE0-D3//GFN2-xTB overestimates reaction barriers on average compared to DLPNO-CCSD(T)//PBE-D3.

To bring these errors into perspective, we extracted the reaction barriers for PBE0-D3//PBE-D3 and PBE-D3//PBE-D3 from the DLPNO-CCSD(T)//PBE-D3 exploration and plotted their $\Delta\Delta G_I^\ddagger$ in Figs.~\ref{fig:gfn2-error-plots}(c) and \ref{fig:gfn2-error-plots}(d), respectively. PBE0-D3//PBE-D3 shows a MAD of $10.3~\si{kJ.mol^{-1}}$ that is lower than for PBE0-D3//GFN2-xTB ($15.1~\si{kJ.mol^{-1}}$) and a mean error of $9.1~\si{kJ.mol^{-1}}$. This implies that PBE0-D3//PBE-D3 underestimates the reaction barrier on average. Note that PBE0-D3//GFN2-xTB overestimates it, even though the electronic energy contributions are calculated with the same DFT method. Moreover, the importance of the DFT functional with which the electronic energies are calculated becomes evident for PBE-D3//PBE-D3. PBE-D3//PBE-D3 has a high MAD of $32.0~\si{kJ.mol^{-1}}$ because it systematically underestimates the reaction barrier, as is shown by the high mean of $\Delta\Delta G_I^\ddagger$ which is $29.8~\si{kJ.mol^{-1}}$.

The free energy differences $\Delta G_n$ [Fig.~\ref{fig:gfn2-error-plots}(b)] show a striped pattern as a function of $G_n(\text{PBE0-D3//PBE-D3})$ because of the high absolute values of the total absolute energies $G_n(\text{PBE0-D3//PBE-D3})$. Furthermore, the $\Delta G_n$ increases on average with decreasing $G_n(\text{PBE0-D3//PBE-D3})$. The mean and MAD of $\Delta G_n$ are $12.0~\si{kJ.mol^{-1}}$ and $15.8~\si{kJ.mol^{-1}}$, respectively, reflecting the anti-correlation between $G_n(\text{PBE0-D3//PBE-D3})$ and $\Delta G_n$.

To be consistent with the MAD of $\Delta\Delta G_I^\ddagger$, we chose the uncertainty of $\Delta G^\ddagger_I$ with PBE0-D3//GFN2-xTB to be a constant value of $u(\Delta G^\ddagger_I) = 15.0~\si{kJ.mol^{-1}}$. Furthermore, we chose the uncertainty bounds for $G_n$ and PBE0-D3//GFN2-xTB as $u(G_n) = 10.0~\si{kJ.mol^{-1}}$, as a compromise between the MAD and the fact that $\Delta G_n$ is significantly smaller for small molecules.

Even for our most accurate electronic structure model combination DLPNO-CCSD(T)//PBE-D3, there remain a large number of error sources, such as the approximations intrinsic to local coupled cluster, errors in the solvation-free energy approximation, anharmonicities in the vibrations, and significant contributions from the conformational entropy, which all contribute to the uncertainty of $\Delta G^\ddagger_I$. Quantifying all these uncertainty sources would be highly desirable but exceeds the scope of this work. Therefore, we restricted our investigation to the uncertainty of the approximations from the DLPNO ansatz by calculating $\Delta G^\ddagger_I$ with normal (pair truncation threshold $t_\mathrm{pair} = 1\cdot 10^{-4}~\si{E_h}$, PNO truncation threshold $t_\mathrm{PNO} = 3.33\cdot 10^{-7}$) and tight PNO ($t_\mathrm{pair} = 1\cdot 10^{-5}~\si{E_h}$, $t_\mathrm{PNO} = 1\cdot 10^{-7}$) settings and taking the absolute differences $\delta_\mathrm{PNO} \Delta G^\ddagger_I$. Accuracies for relative energies of $1~\si{kcal/mol}$ and $1~\si{kJ/mol}$ were reported previously for normal and tight PNO settings, respectively, compared to canonical CCSD(T)\cite{Liakos2015}. We defined the uncertainty as
\begin{align}
    u(\Delta G^\ddagger_I) = \min\left( 5.0~\si{kJ.mol^{-1}}, \delta_\mathrm{PNO} \Delta G^\ddagger_I \right)~.
\end{align}
We chose a minimum uncertainty of $5.0~\si{kJ.mol^{-1}}$ to account for the other error sources that we did not quantify in this work.

Because $G_n$ are absolute energies in our model, there is no clear approach to quantify the uncertainty in the electronic energy contribution from PBE0-D3 in the PBE0-D3//PBE-D3 method combination. Apart from the electronic energy uncertainty, the same uncertainty sources are present for DLPNO-CCSD(T)//PBE-D3. Therefore, we chose a constant uncertainty of $5.0~\si{kJ.mol^{-1}}$. An overview of our uncertainty estimates is given in Table~\ref{tab:uq}.

\begin{table}
    \centering
    \caption{Uncertainty estimates for the local OAT and Morris' sensitivity analysis, in $\si{kJ.mol^{-1}}$.}
    \begin{tabular}{l|c c}
    \toprule\toprule
         & $u(G_n)$ & $u(\Delta G^{\ddagger}_I)$ \\ \midrule
        PBE0-D3//GFN2-xTB & $10.0$ & $15.0$\\
        PBE0-D3//PBE-D3 & $5.0$ & -- \\
        DLPNO-CCSD(T)//PBE-D3 & -- & $\min\left( 5.0~\si{kJ.mol^{-1}}, \delta_\mathrm{PNO} \Delta G^\ddagger_I \right)$ \\
        \bottomrule\bottomrule
    \end{tabular}
    \label{tab:uq}
\end{table}

\section{Results}\label{sec:results}
\subsection*{Local Sensitivity Analysis}

\begin{figure}
    \centering
    \includegraphics[width=\textwidth]{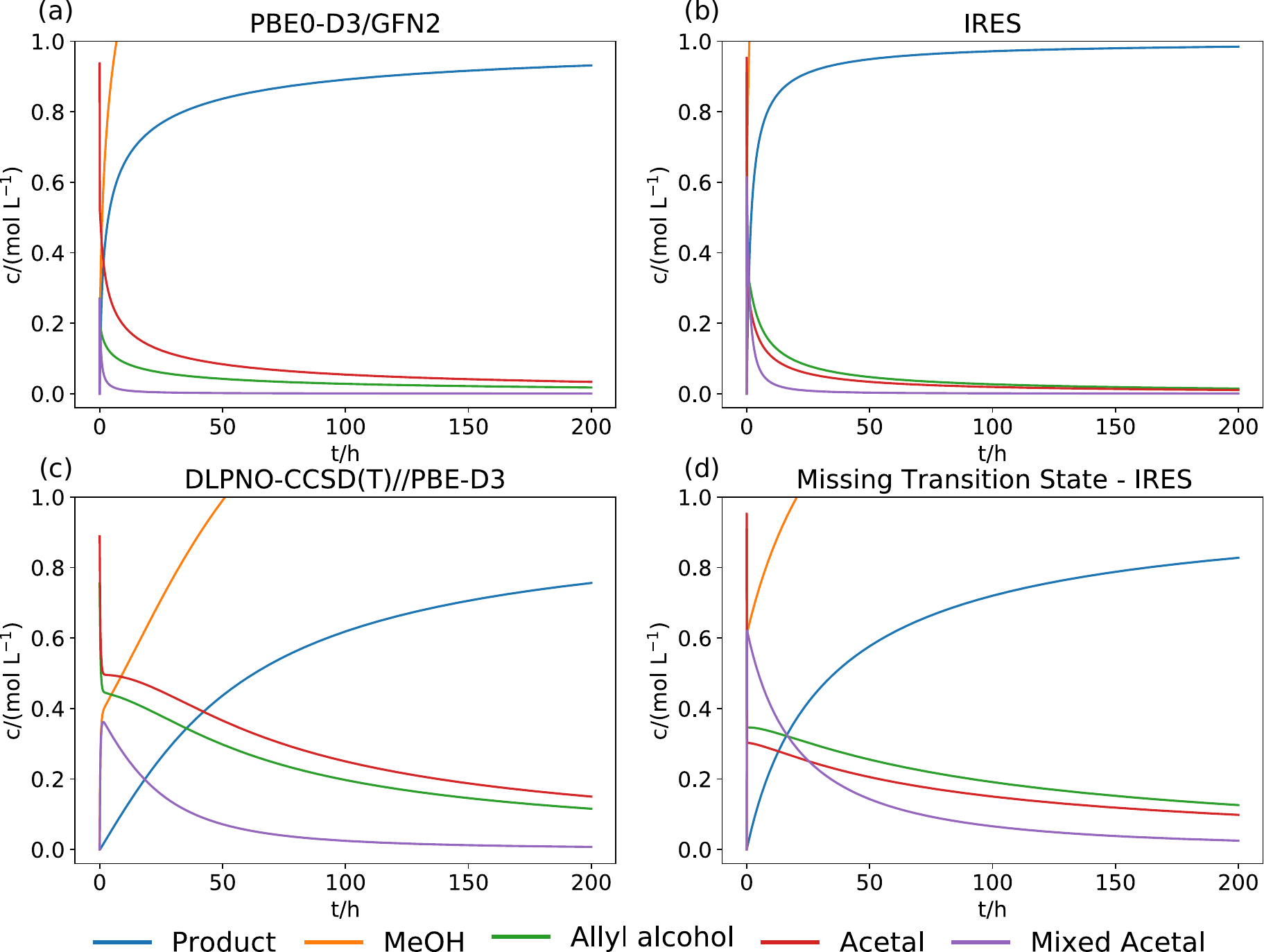}
    \caption{Concentration trajectories of reactants, products, and main intermediates for the Eschenmoser--Claisen allyl rearrangement of the allyl alcohol. The trajectories were calculated based on the reaction networks explored with PBE0-D3//GFN2-xTB (a), the local OAT sensitivity-based IRES exploration (b), DLPNO-CCSD(T)//PBE-D3 (c), and the IRES-based exploration without a favorable transition state for the MeOH catalyzed MeOH elimination from \textbf{a}.}
    \label{fig:allyl-alcohol-comparison}
\end{figure}

To analyze the efficiency of the local OAT sensitivities-based IRES exploration for the Eschenmoser--Claisen rearrangement of allyl alcohol, we compared the microkinetic model extracted from the IRES exploration to the models obtained from the PBE0-D3//GFN2-xTB and DLPNO-CCSD(T)//PBE-D3 explorations. The concentration trajectories of the main product \textbf{a4}, methanol, the allyl alcohol \textbf{a1}, N, N-dimethylacetamide-dimethyl acetal \textbf{a}, and the mixed acetal \textbf{a2} (sum of the concentrations for both enantiomers) are shown in Fig.~\ref{fig:allyl-alcohol-comparison} (a)--(c).

The microkinetic model extracted from the reaction network explored with our IRES-based approach [Fig.~\ref{fig:allyl-alcohol-comparison}(b)] shows the fastest product formation, reaching a concentration of more than $0.9~\si{mol.L^{-1}}$ within $24~\si{h}$. The product formation predicted by the PBE0-D3//GFN2-xTB model [Fig.~\ref{fig:allyl-alcohol-comparison}(a)] is slower, showing only $0.76~\si{mol.L^{-1}}$ after $24~\si{h}$, while the product concentration predicted by the model based on the DLPNO-CCSD(T)//PBE-D3 [Fig.~\ref{fig:allyl-alcohol-comparison}(c)] is only $0.25~\si{mol.L^{-1}}$ after $24~\si{h}$, and therefore significantly slower than both other models.

\begin{figure}
    \centering
    \includegraphics[width=0.7\textwidth]{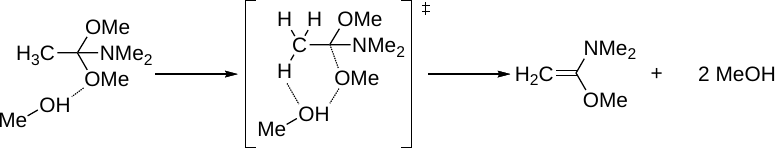}
    \caption{Mechanistic sketch of the methanol-catalyzed methanol elimination from \textbf{a}.}
    \label{fig:meoh-elimination}
\end{figure}

The disagreement between DLPNO-CCSD(T)//PBE-D3 and the IRES-based model is somewhat surprising since the refinement-based approach should systematically improve the parameters from PBE0-D3//GFN2-xTB to DLPNO-CCSD(T)//PBE-D3. The difference between both models is due to the significantly lower free energy of activation of the methanol-catalyzed methanol elimination from the initial acetal \textbf{a} for the IRES-based model compared to DLPNO-CCSD(T)//PBE-D3, shown in Fig.~\ref{fig:meoh-elimination}. To illustrate the effect of this favorable transition state, we removed it from the reaction network. After removing it from the network, the resulting concentration trajectories agree qualitatively with the DLPNO-CCSD(T)//PBE-D3 concentrations, as shown in Fig.~\ref{fig:allyl-alcohol-comparison}(d).
Because the lower reaction barrier for the methanol-catalyzed methanol elimination is a result of the refinement with the DLPNO-CCSD(T)//PBE-D3 model combination, the refined reaction network [concentration plots in Fig.~\ref{fig:allyl-alcohol-comparison}(b)] is a better model for the reaction than the pure DLPNO-CCSD(T)//PBE-D3 network, which failed to find this transition state. It is likely that the pure DLPNO-CCSD(T)//PBE-D3 did not discover this transition state because it relied exclusively on the Newton-trajectory-type approach to locate transition state guesses. By contrast, the IRES-based strategy employed a double-ended curve optimization to locate transition state guesses for the refinement, which was more successful in this case.

Furthermore, the IRES-based reaction network exploration required significantly fewer high-cost calculations, as shown in Table~\ref{tab:calc_comparison}. While the overall number of reaction trial calculations (single-ended or double-ended transition state searches) with GFN2-xTB for the IRES-based exploration is 29150, and therefore higher than the number of reaction trial calculations required for the pure DLPNO-CCSD(T)//PBE-D3 exploration (19763 trials), these calculations are multiple orders of magnitude faster and contribute only little to the required computational resources. 
Compared to the high number of 19763 PBE-D3-based exploration trials for the  DLPNO-CCSD(T)//PBE-D3 exploration, only 199 PBE-D3-based trials were needed for the IRES-based exploration, reducing computational demands by nearly a factor of 100. These savings by two orders of magnitude are significantly higher than the computational time spent on less demanding additional 434 PBE-D3 structure optimization for the $G_n$ refinement. The structure optimizations require only few computational resources compared to an exploration trial calculation because each reaction trial calculation consists of several structure optimizations, a transition state search, and intrinsic reaction coordinate scans\cite{Unsleber2022}.

\begin{table}
    \centering
    \caption{Number of DFT and GFN2-xTB reaction trial calculations, DLPNO-CCSD(T) single point calculations (sp.), and DFT geometry optimizations (opt.) required for the DLPNO-CCSD(T)//PBE-D3 exploration and the IRES based on local OAT sensitivities.}
    \begin{tabular}{l| c c c}
    \toprule\toprule
         & DLPNO-CCSD(T)//PBE-D3 & local OAT IRES  \\ \midrule
        DFT Trials & 19763 & 199\\ 
        GFN2-xTB Trials & -- & 29150\\
        DLPNO-CCSD(T) sp. & 2561 & 468 \\ 
        DFT opt. & -- & 434 \\
        \bottomrule\bottomrule
    \end{tabular}
    \label{tab:calc_comparison}
\end{table}

The computational savings are smaller for the DLPNO-CCSD(T) single-point calculations because, in the DLPNO-CCSD(T)//PBE-D3 exploration, electronic energies were only refined for elementary steps with a barrier lower than $250.0~\si{kJ.mol^{-1}}$s. Nevertheless, the IRES-based exploration required more than a factor 5 fewer DLPNO-CCSD(T) calculations than the full DLPNO-CCSD(T)//PBE-D3 exploration (468 vs. 2561 calculations).

\begin{figure}
    \centering
    \includegraphics[width=\textwidth]{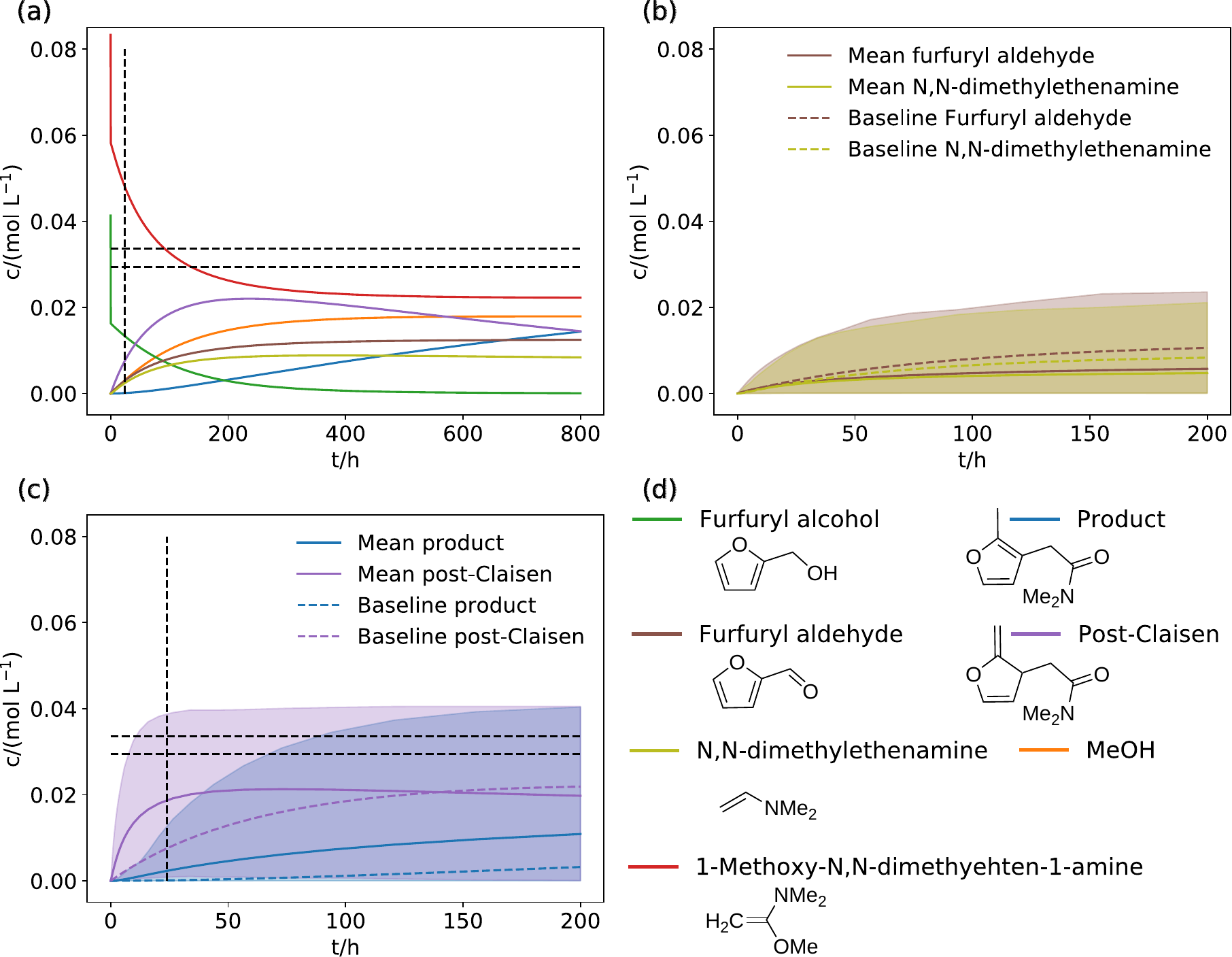}
    \caption{Concentration trajectories simulated for the reaction network explored with the local OAT sensitivity-based IRES. (a) Concentration trajectories for the most populated compounds were calculated with the best available parameters (baseline). (b, c) Uncertainty estimation based on the Morris sensitivity analysis model evaluations. $90~\%$ of trajectories are within the shaded area. ``Baseline'' and ``Mean'' denote the trajectory calculated with the baseline (best available) parameters and the mean of the simulation ensemble, respectively. (d) Compound Lewis structures and trajectory color coding. The black dashed lines denote the experimental yield of $70~\%$-$80~\%$ after $24~\si{h}$.}
    \label{fig:furfuryl-comparison}
\end{figure}

The concentration trajectories calculated with the microkinetic modeling parameters from the local OAT sensitivity-based IRES exploration of the Eschenmoser--Claisen rearrangement of furfuryl alcohol are shown in Fig.~\ref{fig:furfuryl-comparison} (a). The microkinetic model predicts only very slow product formation. Most of the reactants are converted to the post-Claisen compound \textbf{f4} only, and significant concentrations of furfuryl aldehyde and N,N-dimethyletheneamine (see Fig.~\ref{fig:furfuryl-comparison}(d) for the Lewis structures) are produced, effectively leading to a deactivation of the reactants.
However, for this reaction, the experimental yield after $24~\si{h}$ starting from $42~\si{mmol}$ furfuryl alcohol and $84~\si{mmol}$ 1-methoxy-N,N-dimethylethen-1-amine was reported to be $70~\%$-$80~\%$\cite{Felix1969}. This experimental observation suggests that the free energy of activation for the rearomatization (\textbf{f4} $\rightarrow$ \textbf{f5}) of the post-Claisen compound \textbf{f4} is overestimated, and that \textbf{f4} is formed too slowly in our model.

To better understand the disagreement of our model with the experimental observation and to estimate the uncertainty in the concentrations, we performed a Morris sensitivity analysis with the same settings discussed in Section~\ref{sec:sensitivity-analysis}. The mean concentrations of all model evaluations, the $90~\%$ percentiles, and the concentration trajectories calculated with the baseline (best) parameters are shown for the post-Claisen compound \textbf{f4} and the product \textbf{f5} in Fig.~\ref{fig:furfuryl-comparison}(b), and for the side-products furfuryl aldehyde and N,N-dimethyletheneaminin in Fig.~\ref{fig:furfuryl-comparison}(c).
The mean concentrations predicted for the product \textbf{f5} and the post-Claisen intermediate \textbf{f4} are significantly higher than the concentrations predicted by the baseline model.
This clearly shows that a faster formation of the post-Claisen intermediate \textbf{f4} and the product \textbf{f5} is possible within the uncertainty assumed for the microkinetic modeling parameters. However, the experimental yields are not covered by the $90~\%$ percentile of the product \textbf{f5}, suggesting that we may have underestimated the error in the parameters.

The side-products furfuryl aldehyde and N,N-dimethyletheneaminin remain at moderate concentrations even if we consider their concentration's uncertainty [see Fig.~\ref{fig:furfuryl-comparison}(c)]. Therefore, our model is qualitatively correct as it predicts the experimental product \textbf{f5} and the post-Claisen intermediate \textbf{f4} as the main reaction products. 

\subsection*{Uncertainty-aware Explorations}

\begin{figure}
    \centering
    \includegraphics[width=\textwidth]{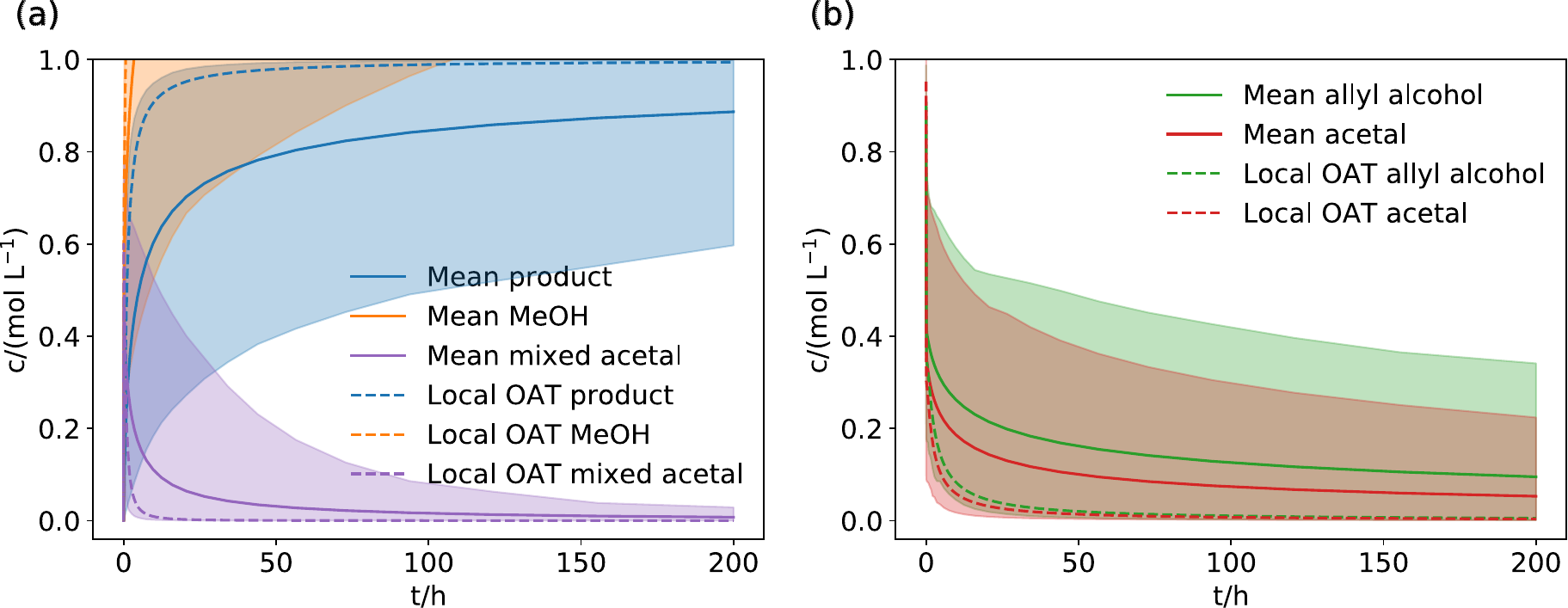}
    \caption{(a) Concentration trajectories for the main products (\textbf{a4}, MeOH), intermediates (\textbf{a2}), and (b) reactants (\textbf{a}, \textbf{a1}) calculated based on the reaction network of the Eschenmoser--Claisen rearrangement of allyl alcohol with the uncertainty-aware exploration approach. $90~\%$ of trajectories are within the shaded area. ``Mean'' denotes the mean trajectory of the simulation ensemble and ``Local OAT'' denotes the trajectories from the exploration based on local OAT sensitivities.}
    \label{fig:allyl-alcohol-morris}
\end{figure}

The mean concentration trajectories for the product \textbf{a4} of the rearrangement of allyl alcohol, methanol, the mixed acetal \textbf{a2}, and the reactants \textbf{a}/\textbf{a1} calculated with our uncertainty-aware exploration approach are shown with their counterpart from the local OAT-based exploration in Fig.~\ref{fig:allyl-alcohol-morris}. The mean trajectories show slower formation of the product \textbf{a4} and, in turn, slower reactant consumption than the results from the local OAT-based exploration. However, in all cases, the local OAT-based concentration trajectories are within the $90~\%$ percentiles of the uncertainty-aware exploration trajectories, \emph{i.e.}, the uncertainty-aware exploration and local OAT-based exploration agree in their predictions.

\begin{figure}
    \centering
    \includegraphics[width=\textwidth]{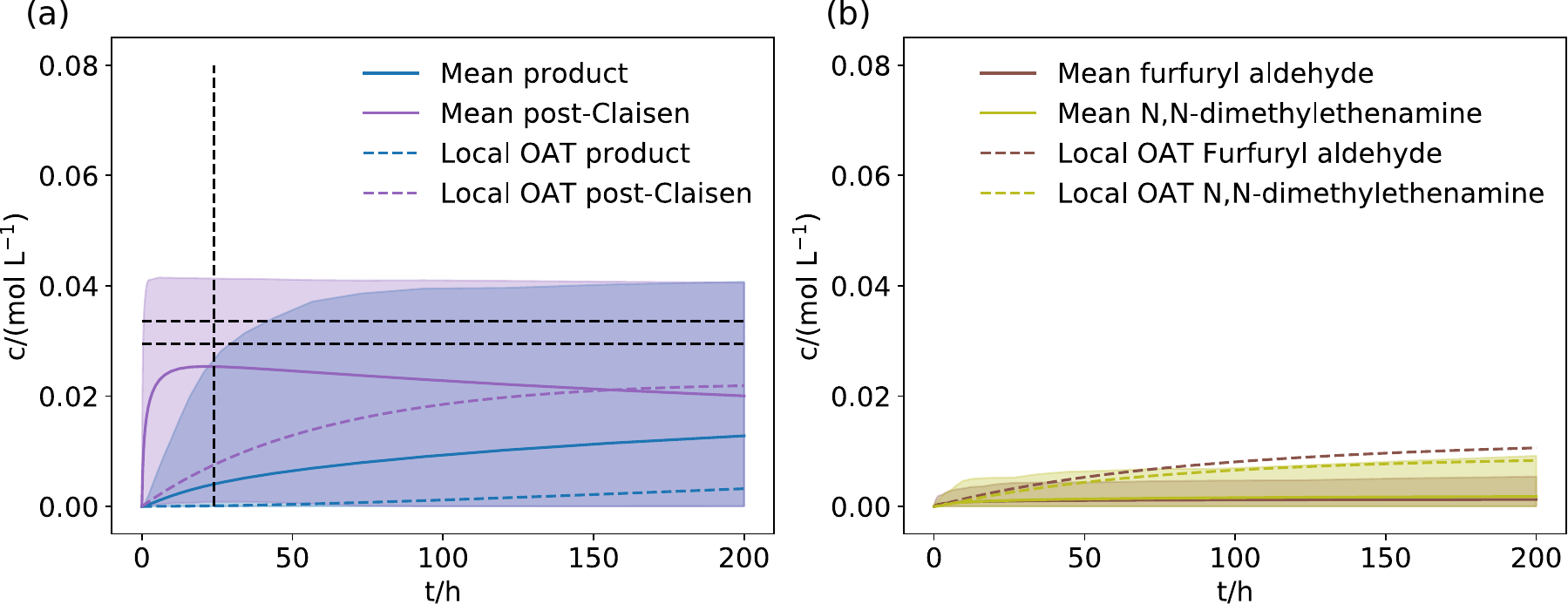}
    \caption{Concentration trajectories for the main product \textbf{f5}, intermediate \textbf{f4} (a), and side products (b) calculated based on the reaction network of the Eschenmoser--Claisen rearrangement of furfuryl alcohol with the uncertainty-aware exploration approach. $90~\%$ of trajectories are within the shaded area. ``Mean'' denotes the mean trajectory of the simulation ensemble and ``Local OAT'' denotes the trajectories from the exploration based on local OAT sensitivities.}
    \label{fig:furfuryl-morris}
\end{figure}

The concentration trajectories of the product \textbf{f5}, intermediate \textbf{f4}, and the side products furfuryl aldehyde and N,N-dimethylethenamine for the uncertainty-aware exploration of the Eschenmoser--Claisen rearrangement of furfuryl alcohol are shown in Fig.~\ref{fig:furfuryl-morris}. The concentration trajectories for \textbf{f4} and \textbf{f5} are similar to the trajectories calculated based on the Morris sensitivity analysis of the local OAT-based exploration presented in Fig.~\ref{fig:furfuryl-comparison}. The main difference is that the minimum concentration of \textbf{f5} based on the experimental estimate ($70~\%$-$80~\%$ of the reactants), highlighted by the dashed lines in Fig.~\ref{fig:allyl-alcohol-morris}(a), is nearly within the uncertainty estimate for the product shown as the shaded area. By contrast, the experimental estimate is far off the uncertainty shown for the local OAT-based exploration shown in Fig.~\ref{fig:furfuryl-comparison}(b), suggesting that the uncertainty-aware exploration provided a more accurate picture of the reaction kinetics by more reliably identifying and refining the critical reaction channels during the rolling exploration.

Furthermore, the uncertainty-aware exploration predicts that the side-products furfuryl aldehyde and N,N-dimethyletheneamine are essentially not relevant for the reaction model, as shown in Fig.~\ref{fig:furfuryl-morris}(b). Both compounds do not reach significant concentrations within our uncertainty estimates.

An overview of the number of compounds, flasks, and reactions in the final reaction networks of the local OAT-based exploration and the uncertainty-aware exploration, together with the number of reaction trial calculations and refinement calculations needed for the exploration, are given in Tab.~\ref{tab:network-overview}. Even though the KIEA exploration thresholds $\tau_\mathrm{flux}$ and $\tau_\mathrm{max}$ are chosen higher for the uncertainty-aware exploration of the rearrangement of allyl alcohol, the uncertainty-aware exploration uncovers nearly 1000 additional compounds (1622 vs. 2621), 547 more flasks (1073 vs. 1630), and 1919 more reactions (3200 vs. 5119) compared to the local OAT-based exploration. This significant increase in discovered compounds and flasks can be attributed to the increased number of bimolecular reaction trials, which were 28486 for the local OAT-based exploration and 42011 for the uncertainty-aware exploration. However, most of the newly found compounds, flasks, and reactions did not contribute significantly to the uncertainty of the concentration prediction and were, therefore, not refined, as is evident from the only moderately increased number of elementary step refinement calculations between the explorations (local OAT-based 199, uncertainty-aware 256). Furthermore, the only compound explored in addition to the uncertainty-aware exploration was propanol, originating from the disproportionation of allyl alcohol into propanol and prop-2-en-1-al, which did not reach any significant concentrations even if the uncertainty was considered.

\begin{table}
    \centering
    \caption{Overview of the number of compounds, flasks, and reactions in the networks, the number of unimolecular and bimolecular single-ended reaction trials, and double-ended refinement calculations required to explore the networks. The numbers in parenthesis denote the number of flasks/compounds fulfilling the exploration criteria of KIEA. Databases containing the full reaction networks are available on Zenodo~\cite{Bensberg2023i, Bensberg2023j}.}
    \begin{tabular}{l|c c c c c c}
    \toprule\toprule
                          & \multicolumn{6}{c}{Allyl alcohol} \\ \midrule
                          & Compounds & Flasks & Reactions & Unimol.& Bimol. & Refin. \\ \midrule
        local OAT IRES    & 1622 (14) & 1073 (31) & 3200 & 644 & 28486 & 199\\
        uncertainty-aware & 2621 (15) & 1630 (32) & 5119 & 739 & 42011 & 256\\ \midrule
                          & \multicolumn{6}{c}{Furfuryl alcohol} \\ \midrule
        local OAT IRES    & 13644 (22) &  7696 (40) & 24195 & 9806 & 163354 & 218\\
        uncertainty-aware & 13277 (25) & 7270 (45) & 23077 & 8538 & 153192 & 248\\
        \bottomrule\bottomrule
    \end{tabular}
    \label{tab:network-overview}
\end{table}

The number of exploration trials and refinement calculations required to converge the exploration of the Eschenmoser--Claisen rearrangement of furfuryl alcohol are very similar between the uncertainty-aware and the local OAT-based approaches. The uncertainty-aware exploration ansatz required roughly 10,000 fewer bimolecular reaction trial calculations (163,354 vs. 153,192) and 1,268 fewer unimolecular reaction trial calculations, while the number of double-ended elementary step refinement calculations increased by $30$ from $218$ to $248$.

\section{Conclusions\label{sec:conclusion}}
We presented a fully automated first-principles exploration approach, KIEA-IRES, that combines automated reaction network exploration, microkinetic modeling-based exploration steering, sensitivity analysis, and refinement of kinetic parameters for reactions, compounds, and flasks.

We explored the reaction network of the Eschenmoser--Claisen rearrangement containing tens of thousands of reactions and compounds with KIEA-IRES. KIEA-IRES correctly predicted the product of the rearrangement of furfuryl alcohol known from experiment and predicted the product of the rearrangement of allyl alcohol (not reported experimentally so far), as expected based on experimental studies for similar molecules\cite{Wick1964}.

The exploration approach requires no prior knowledge of the chemistry that is explored. The only remaining input of general chemistry knowledge in our approach is the restriction of the reaction trial calculations by a small set of rules applicable to organic chemistry, as discussed in Section~\ref{sec:comp-methods}. These rules could be replaced in the future by ’first-principles heuristics’\cite{Bergeler2015} based on the analysis of partial charges, Fukui functions, or other concepts\cite{Bergeler2015, Grimmel2019, Grimmel2021}.

Our approach effectively exploits the fact that, out of the thousands of reactions and compounds in the network, only a small subset determines the kinetics. These reactions and compounds were automatically identified by global or local sensitivity analysis. The kinetic parameters encoded for them were refined with more accurate but computationally more costly quantum chemical methods. This refinement-driven approach led to significant computational savings compared to a full exploration with accurate but costly methods without loss in accuracy. For instance, IRES-KIEA required almost a factor of 100 fewer computationally costly DFT-based exploration trial calculations than a full DLPNO-CCSD(T)//PBE-D3-based KIEA reference exploration.

Furthermore, we compared the activation energies and free energies calculated for GFN2-xTB structures with the same quantities calculated for PBE-D3 structures. We found a significant spread in the error for the activation energies and a correlation between the error in the free energies and the absolute free energy value. The large spread for the activation energies highlights the importance of considering the uncertainty in the kinetic parameters in microkinetic modeling simulations and even in qualitative discussions of reaction mechanisms based on activation energies.

Our local OAT-based explorations and the uncertainty-aware exploration protocol relying on Morris sensitivity analysis both predicted the same products and kinetics for the example reactions. Nevertheless, the uncertainty-aware exploration approach is conceptually more appealing since it directly provides meaningful uncertainties for the concentrations and considers the microkinetic modeling parameters as distributions rather than as fixed values, which may prove crucial if the initial exploration method (here PBE0-D3//GFN2-xTB) turns out to be qualitatively wrong by favoring an incorrect reaction path.
The local OAT-based sensitivity analysis can be considered a low-cost alternative for reaction networks in which the microkinetic model is extraordinarily large and the flux-based screening procedure cannot reduce the number of model parameters.

\section*{Data Availability}
The databases containing all information to reproduce this study are provided on \textsc{Zenodo}~\cite{Bensberg2023i, Bensberg2023j}.

\providecommand{\latin}[1]{#1}
\makeatletter
\providecommand{\doi}
  {\begingroup\let\do\@makeother\dospecials
  \catcode`\{=1 \catcode`\}=2 \doi@aux}
\providecommand{\doi@aux}[1]{\endgroup\texttt{#1}}
\makeatother
\providecommand*\mcitethebibliography{\thebibliography}
\csname @ifundefined\endcsname{endmcitethebibliography}
  {\let\endmcitethebibliography\endthebibliography}{}

\end{document}